\documentclass[twocolumn,10pt]{asme2e}

 \pdfoutput=1
\usepackage{graphicx}
\usepackage{caption}
\usepackage{subcaption}
\usepackage{lipsum}
\usepackage{amsmath}
\usepackage{multicol}

\title{A New RANS Correction to Account for Varying Viscosity Effects}

\author{Victor Coppo Leite
    \affiliation{
	Ken and Mary Alice Lindquist\\Department of Nuclear Engineering\\
	Pennsylvania State University\\
	University Park, PA 16802\\
    Email: vbc5085@psu.edu
    }
}

\author{Elia Merzari
    \affiliation{
	Ken and Mary Alice Lindquist\\Department of Nuclear Engineering\\
	Pennsylvania State University\\
	University Park, PA 16802\\
    Email: ebm5153@psu.edu
    }
}

\begin{document}

\maketitle

\begin{abstract}
{It has previously been shown that by increasing the Reynolds number across a channel by spatially varying the viscosity does not cause an immediate change in the size of turbulent structures and a delay is in fact observed in both wall shear and friction Reynolds number (Coppo Leite, V, \& Merzari, E., Proceedings of the ASME 2020 FEDSM, p. V003T05A019). Furthermore, it is also shown that depending on the length in which the flow condition changes, turbulence bursts are observed in the turbulence field. For the present work we propose a new version of the standard Reynolds Averaged Navier Stokes (RANS) \(k-\tau\) model that includes some modifications in the production term in order to account for these effects. The new proposed model may be useful for many engineering applications as turbulent flows featuring temperature gradients and high heat transfer rates are often seen in heat exchangers, combustion chambers and nuclear reactors. In these applications, thermal and viscous properties of the working fluid are important design parameters that depend on temperature; hence it is likely to observe strong gradients on these scalars’ fields. To accomplish our goal, the modifications for the \(k-\tau\) model are implemented and tested for a channel flow with spatial varying viscosity in the streamwise direction. The numerical simulations are performed using Nek5000, a spectral-element code developed at Argonne National Laboratory (ANL). Finally, the results considering a turbulence channel using the proposed model are compared against data obtained using Direct Numerical Simulations from the earlier work.}
\end{abstract}

\begin{nomenclature}
\entry{$a$}{Viscosity linear coefficient}
\entry{$C$}{Contraction parameter}
\entry{$Re_\tau^* (x)$}{Convolution function}
\entry{$k$}{Turbulent kinetic energy}
\entry{$P_k,P_\tau$}{Production term for $k$ and $\tau$}
\entry{$P_\kappa$}{Corrected production term for $k$}
\entry{$R$}{Relaxation parameter}
\entry{$Re$}{Reynolds number}
\entry{$Re_{\tau}$}{Friction Reynolds number}
\entry{$u_{\tau}$}{Friction velocity}
\entry{$\overline{u_iu_j}$}{Reynolds-stress tensor}
\entry{$x$}{Streamwise direction}
\entry{$y$}{Wall-normal direction}
\entry{$\alpha_\omega$}{Coefficient of production term of $\omega$}
\entry{$\beta_k,\beta_\omega$}{Coefficient of dissipation terms of $k$ and $\omega$}
\entry{$\delta$}{Height of the turbulence channel}
\entry{${\delta}_{i}$}{Streamwise position where Region $i$ starts, ${i}$=I,II and III}
\entry{${\Lambda}$}{Signal contribution in the delayed function}
\entry{$\mu_k,\mu_\omega$}{Viscosity coefficient of $k$ and $\omega$}
\entry{$\mu,\mu_t$}{Dynamic molecular and eddy viscosity coefficients}
\entry{$\sigma_k,\sigma_\omega,\sigma_d$}{Coefficients of diffusion terms of $k$ and $\omega$}
\entry{$\nu$}{Kinematic Viscosity}
\entry{$\tau$}{Turbulent time scale $\tau=1/\omega$}
\entry{$\omega$}{Specific turbulent dissipation rate}

\entry{${\tau}_w$}{Wall Shear stress }

\end{nomenclature}

\section*{INTRODUCTION}

In the present paper, we propose some modifications on the standard \(k-\tau\) Reynolds Averaged Navier Stokes (RANS) model from Ref.~\cite{kalitzin1996}. The modifications are made in order to account for the effects caused by spatially varying viscosity observed in an earlier studies \cite{leite2020,leite2021}. In these studies, we performed Direct Numerical Simulations (DNS) of a channel flow where viscosity is imposed to change spatially throughout the streamwise direction whereas the flow is homogeneous in the spanwise direction. The case set up is a simplification of a heated channel, since the temperature field has not been evaluated. Nevertheless, the proposed set up is valuable for future works considering variable-viscosity effects, with relevance for cooling and heating applications.

For the present work we consider the same configuration for the channel flow from Ref.~\cite{leite2021}. The channel is extended in the streamwise direction where the Reynolds number linearly increase by locally varying the viscosity within a ramp region. The channel features three regions: in Region I, viscosity remains constant and the flow is at a Reynolds number of \(Re=5,000\); in Region II, i.e. the ramp region, viscosity decreases by the inverse of the streamwise length causing the Reynolds number to increase linearly from its former value up to \(Re=10,000\); and in Region III, we impose viscosity to remain constant again but now the flow is at \(Re=10,000\). It should be noted that since the flow is homogeneous in the spanwise direction and only simulations using RANS models are performed in the present work, we further simplified the models by only considering two dimensions, i.e. the vertical and the streamwise directions.

In Ref.~\cite{leite2021}, different effects in the turbulence field due to the imposition of the ramp region are reported. Part of that investigation consisted on developing a convolution function that served as a tool to model the spatial delay from the DNS results. For the present work, we use that function to adjust the standard \(k-\tau\) formulation such that the effects observed in the DNS are retrieved by the RANS model. Namely, the production term for the turbulent kinetic energy (TKE) from the standard \(k-\tau\) model has been modified by a correction factor computed that is based on the convolution function obtained using the DNS data.

Additionally, in Ref.~\cite{leite2021} it is explained that these effects are essentially caused by turbulent structures dynamics, more specifically the behavior of the low-speed streaks (Kline et al.~\cite{kline1967}). Thus, it is unlikely that any of the standard RANS models has the capability to retrieve the physics behind these effects, since they were not developed considering flows undertaking these conditions.

The goal of the present paper is to show that even though the great complexity on the physics behind these effects, they can be modeled through RANS in a reliable and accurate fashion. For that end, we developed the computational fluid dynamics (CFD) models using Nek5000~\cite{nek5000}, a spectral-element code developed at Argonne National Laboratory. Nek5000 is an open source CFD code with excellent scaling capabilities.  Additionally, Nek5000 has some of the most popular two-equations RANS models already implemented, for instance the \(k-\epsilon\), the SST-\(k-\epsilon\), the \(k-\omega\) and the \(k-\tau\) models. Finally, the \(k-\tau\) model is considered in the present study due to three reasons: first, wall functions are not a requirement, it is not as stiff as the \(k-\epsilon\) is in the viscous region and; third, the turbulent time scale \(\tau=1/\omega\) does not exhibit a singular behavior at the wall like the turbulent dissipation does in the \(k-\omega\) model.

The present work is structured as the following: the simulation details are provided in the second section, in the third section the \(k-\tau\) formulation and the proposed modification are presented. Next, in the results section a verification of the model is presented along the results of the proposed model. Finally, the conclusion section provides a summary of the current study and future works proposal.


\section*{METHODS - THE SIMULATION DETAILS}

\begin{figure}[htb] 
\centering
\scalebox{0.4}
{\includegraphics{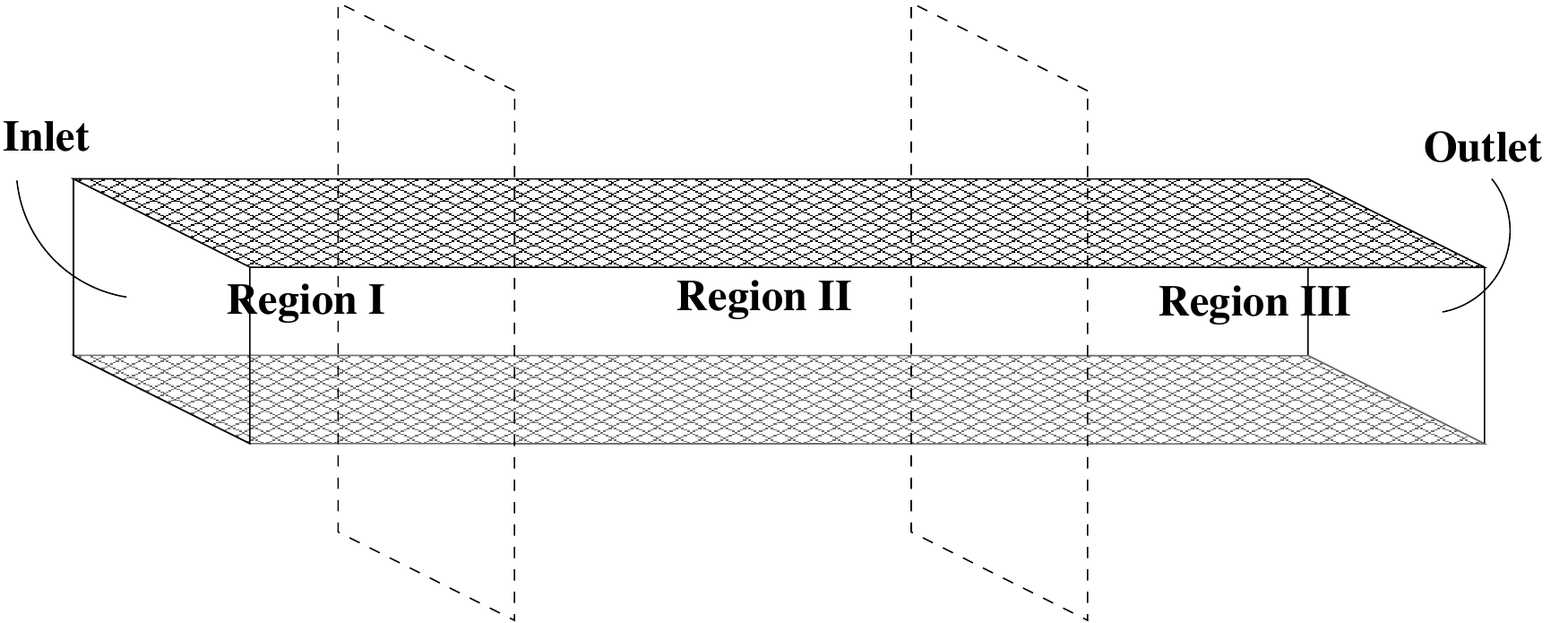}}
\scalebox{0.3}
{\includegraphics{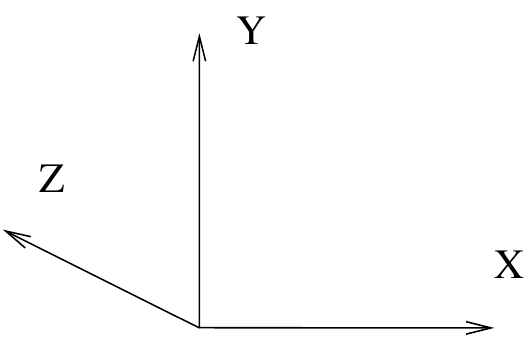}}
\caption{Geometry of the turbulence channel, the channel is divided in three regions where viscosity is imposed to change.}
\label{fig:geometry}
\end{figure}

The CFD simulations performed in the present work were developed using Nek5000~\cite{nek5000}. The test case considered is a channel flow with spatially varying viscosity, schematically represented in Fig.~\ref{fig:geometry}. In the first region of the channel the viscosity is kept at a constant value and the flow is at \(Re=5,000\). In the second region the Reynolds number linearly increase through a ramp as viscosity is imposed to decrease, reaching a final value of \(Re=10,000\) whereas the viscosity changes accordingly. Finally, in the third region, the Reynolds and the viscosity have a constant value again, now equal to those from the end of the second region.

The set-up is similar to the one proposed in Ref.~\cite{leite2020} and \cite{leite2021}. In both those studies, Direct Numerical Simulation (DNS) has been performed in order to study the contribution of the smaller scales to the energy cascade in the turbulence field. Ref.~\cite{leite2020} focused on preliminary DNS results at high Reynolds and \cite{leite2021} focused on detailed DNS results with complete budgets for a lower Reynolds number. For the present work, we refer to the benchmark results from the DNS in order to modify the \(k-\tau\) RANS model already available and tested in Nek5000.

Recently, the more robust unsteady RANS \(k-\tau\) model, which will be further explained in the next section, has been implemented and tested in Nek5000 \cite{nekvandv}. Additionally, Nek5000 features a stress-formulation solver capable of solving all the velocity components at once, which is required for spatially dependent properties.

As a first proof of concept for the modifications to be made in the standard \(k-\tau\) model, we consider reproducing the first case from~\cite{leite2021}. In that case, the flow goes from \(Re=5,000\) (\(Re_\tau=300\)) to \(Re=10,000\) (\(Re_\tau=550\)) through a ramp with length of \(L_{ramp}=0.5\pi\) starting at \(x=5\pi\). Finally, a sufficiently refined mesh was employed to test this case once the \(y^+<1\) criteria is addressed and the model’s results are verified against benchmark data as presented in the Results section.


\section*{METHODS - THE RANS MODEL}

The governing equations for the \(k-\tau\) model implemented in Nek5000 are provided in Ref.~\cite{komega}:

\begin{equation}
\frac{\partial \rho k}{\partial t} + \nabla \cdot (\rho k \vec{u}) =P_k - \beta_k \rho \omega k + \nabla \cdot (\mu_k \nabla k)
\label{eq:kequation}
\end{equation}

\begin{multline}
\frac{\partial \rho \tau}{\partial t} + \nabla \cdot (\rho \tau \vec{u}) = -P_\tau + \beta_\omega \rho
\\ + \nabla \cdot (\mu_\omega \nabla \tau) -8 \mu_\omega \left\Vert \nabla \sqrt{\tau} \right\Vert^2 +C_D
\label{eq:tauequation}
\end{multline}

with \(k\) being the turbulent kinetic energy, \(\rho\) is the density, \(\vec{u}\) is the velocity vector and \(\tau\) is the dissipation time scale. Following, the diffusion coefficient are provided for each of the above mentioned equations:

\begin{equation}
\mu_k=\mu+\sigma_k \mu_t
\label{eq:kdiff}
\end{equation}

\begin{equation}
\mu_\omega=\mu+\sigma_\omega \mu_t
\label{eq:omegadiff}
\end{equation}

with \(\mu\) the molecular-viscosity coefficient and \(\mu_t=\rho k \tau\) is the eddy-viscosity coefficient. The production terms for these equations and the so-called cross diffusion term are given by respectively:

\begin{equation}
P_k= \overline{u_iu_j} \cdot \nabla \vec{u}
\label{eq:kprod}
\end{equation}

\begin{equation}
P_\tau= \frac{\alpha_\omega \tau}{k} P_k
\label{eq:tauprod}
\end{equation}

\begin{equation}
C_D=\sigma_d \rho  \tau \cdot min\{\nabla k \cdot \nabla \tau,0\}
\label{eq:CD}
\end{equation}

where the Reynolds-stress tensor \(\overline{u_iu_j}\) is calculated through the relationship between stress and strain as stated in Ref.~\cite{pope1975}:

\begin{equation}
\overline{u_iu_j} = \frac{2}{3}\delta_{ij}k - \mu_t \left( \frac{\partial u_i}{\partial x_j}+\frac{\partial u_j}{\partial x_i} \right)
\label{eq:reynolds_stress}
\end{equation}

Finally, the model relies on six closure coefficients: \(\alpha_\omega,\beta_k,\beta_\omega\,\sigma_k,\sigma_\omega,\sigma_d\). The values used for these coefficients are the following:

\textit{From Wilcox, in Ref.~\cite{wilcox1993}:}

\begin{equation}
(\alpha_\omega,\beta_k,\beta_\omega)=(0.5,0.09,0.075)
\label{eq:wilcox_coef}
\end{equation}

\textit{From Kok, in Ref.~\cite{kok1999}, the turbulent/non-turbulent (TNT) coefficients:}

\begin{equation}
(\sigma_k,\sigma_\omega,\sigma_d)=(2/3,0.5,0.5)
\label{eq:tnt_coef}
\end{equation}

It should be noted that although one of the most popular two-equations models used is the \(k-\omega\) proposed by Wilcox~\cite{wilcox1993}, particularly because wall-dumping functions are not strictly required, which makes this formulation more attractive for complex geometries. However, it presents a major drawback: the turbulent dissipation has a singular value in the near-wall region once the solution valid in the viscous sublayer is inversely proportional to the square of the wall distance \(y\), i.e. \(\omega \propto 1/y^2\) when \(y \rightarrow 0\)~\cite{wilcox1993}. This matter is avoided by switching the dependent variable to the dissipation time scale \(\tau=1/\omega\). In that case, \(\tau\propto y^2\) near the wall, which grants robustness for the model.

In order to account for the effects due to spatially varying the viscosity, we propose a factor \(C_P\), provided in Eq.~\ref{eq:correction}. This factor is used to correct the production term from Eq.~\ref{eq:kprod}. In the present paper, \(P_\kappa\) is the corrected production term, Eq.~\ref{eq:correctedprod}.

\begin{equation}
C_P=\left( \frac{Re_\tau^* (x)}{Re_\tau (x)} \right)^n
\label{eq:correction}
\end{equation}

\begin{equation}
P_\kappa=C_P P_k
\label{eq:correctedprod}
\end{equation}

In this equation, \(Re_\tau^* (x)\) is computed through Eqs.~\ref{eq:convolution1}-\ref{eq:convolution3}, which is the set of convolution functions developed in Ref.~\cite{leite2021} that served as a tool to model the effects of varying viscosity in the turbulence field throughout Regions I, II and III respectively. The denominator of this ratio is computed with Eq.~\ref{eq:correlation_pope}, which is a correlation valid for fully-developed flows from Ref.~\cite{pope2000}. Finally, \(n=0.5\) for regions where viscosity varies and \(n=3.0\) for regions where viscosity is constant. It should be noted that this last parameter and the correction proposed by Eq.~\ref{eq:correction} were determined through tests.

\begin{equation}
  \begin{gathered}
Re_{\tau,I}^*(x) = 300
\label{eq:convolution1}
  \end{gathered}
\end{equation}

\begin{equation}
  \begin{gathered}
Re_{\tau,II}^*(x) = \frac{a}{R^2}((x-{\delta}_{II})R+e^{-R(x-{\delta}_{II})}-1)+300
\label{eq:convolution2}
  \end{gathered}
\end{equation}

\begin{equation}
  \begin{gathered}
Re_{\tau,III}^*(x) = 300sin\left(\frac{\pi}{P}(x-\sigma_{II})\right)e^{-C(x-\sigma_{II})} \\
+\Lambda(1-e^{-C(x-\sigma_{II})}) + Re_{\tau,II}^*(\sigma_{II})
  \end{gathered}
\label{eq:convolution3}
\end{equation}

Where,

\begin{description}

\item[\(a\)]  The ramp inclination, i.e. \((Re_{\tau,final}-Re_{\tau,initial})/L_{ramp}\)

\item[\(Re_{\tau,i}^*(x)\)]  The delaying function for region $i=I,II$ and $III$

\item[\(\delta_{i}\)] Location in $x$ where Regions $i=II$ and $III$ starts

\item[\(R\)] The relaxation parameter for Region $II$

\item[\(P\)]  The bump width parameter for Region $III$

\item[\(C\)]  The contraction parameter in Region $III$

\item[\(\Lambda\)]  The contribution in \(Re_\tau(x)^*\) from Region $II$ to Region $III$, i.e. \(550-Re_{\tau,II}^*(\delta_{II})\)

\end{description}

\begin{equation}
Re_\tau (x)=0.09\{2Re(x)\} ^{0.88}
\label{eq:correlation_pope}
\end{equation}


\section*{RESULTS}

\subsection*{Model verification}

\begin{figure}[htb]
\centering 
\includegraphics[width=\linewidth]{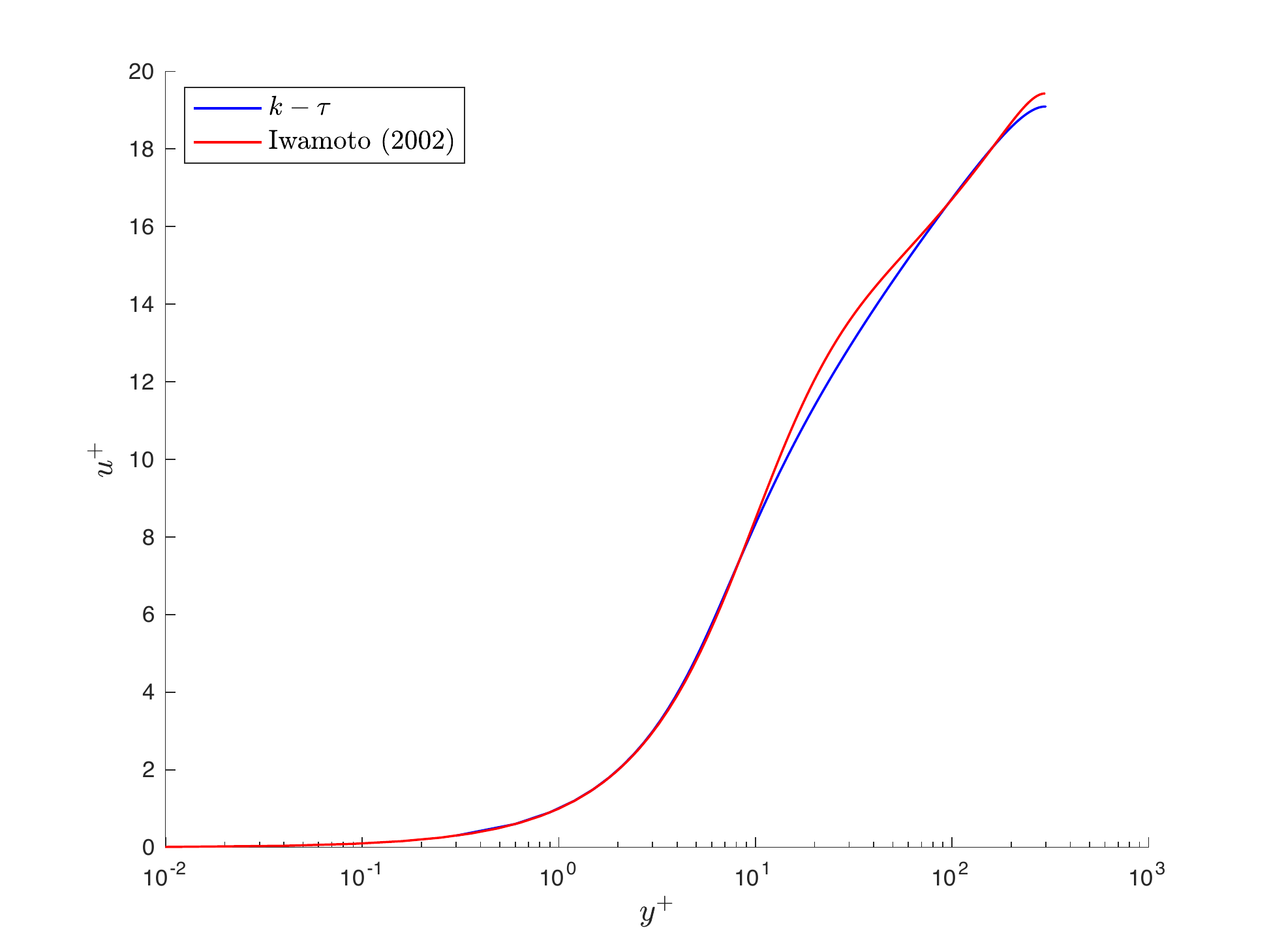}
\caption{Comparing velocity results from the \(k-\tau\) model available in Nek5000 with benchmarked data from Ref.~\cite{iwamoto2002}}
\label{fig:u_verification_inlet}
\end{figure}

\begin{figure}[htb]
\centering 
\includegraphics[width=\linewidth]{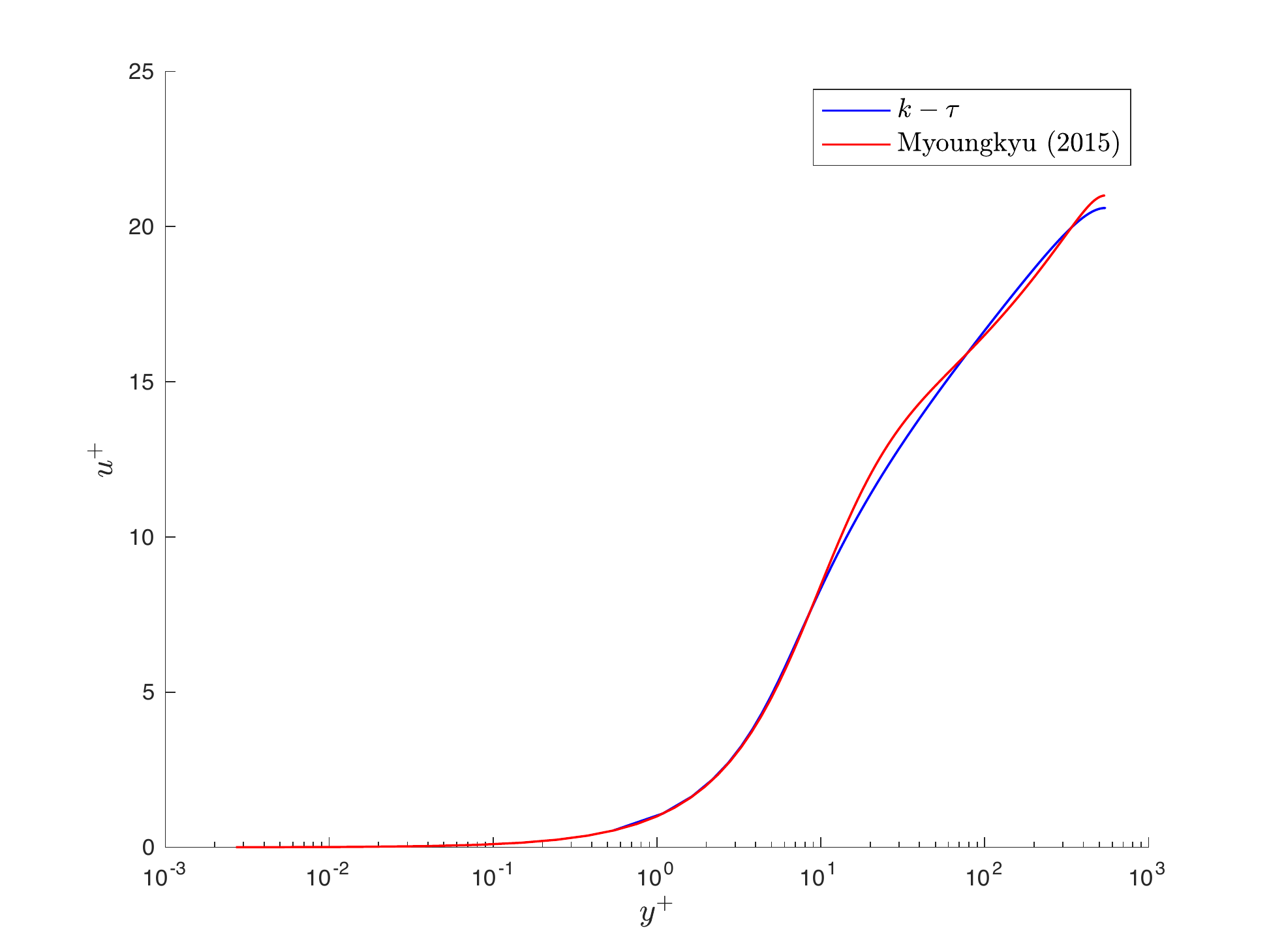}
\caption{Comparing velocity results from the \(k-\tau\) model available in Nek5000 with benchmarked data from Ref.~\cite{myoungkyu2015}}
\label{fig:u_verification_outlet}
\end{figure}

In order to verify the model, Figs.~\ref{fig:u_verification_inlet} and~\ref{fig:u_verification_outlet} shows the streamwise velocity profiles at the inlet and at the outlet repectively of the channel flow. In these plots, the results from the standard \(k-\tau\) model available in Nek5000 are compared with benchmarked data from Refs.~\cite{iwamoto2002} and \cite{myoungkyu2015}, which are valid for flows at $Re_\tau=300$ and at $Re_\tau=550$ respectively. The superscript $+$ indicates that non-dimensional values are plotted. For the $x$ axis the normalization is obtained with the viscous length scale $\delta_\nu=\nu/u_\tau$, where $\nu$ is the viscosity and $u_\tau=\sqrt{\tau_w/\rho}$ is the friction velocity, being $\tau_w$ the mean wall shear stress and $\rho$ the density. For the velocity, the normalization is obtained by means of the friction velocity $u_\tau$.

\subsection*{The proposed RANS model results}

\begin{figure}[htb]
\centering 
\includegraphics[width=\linewidth]{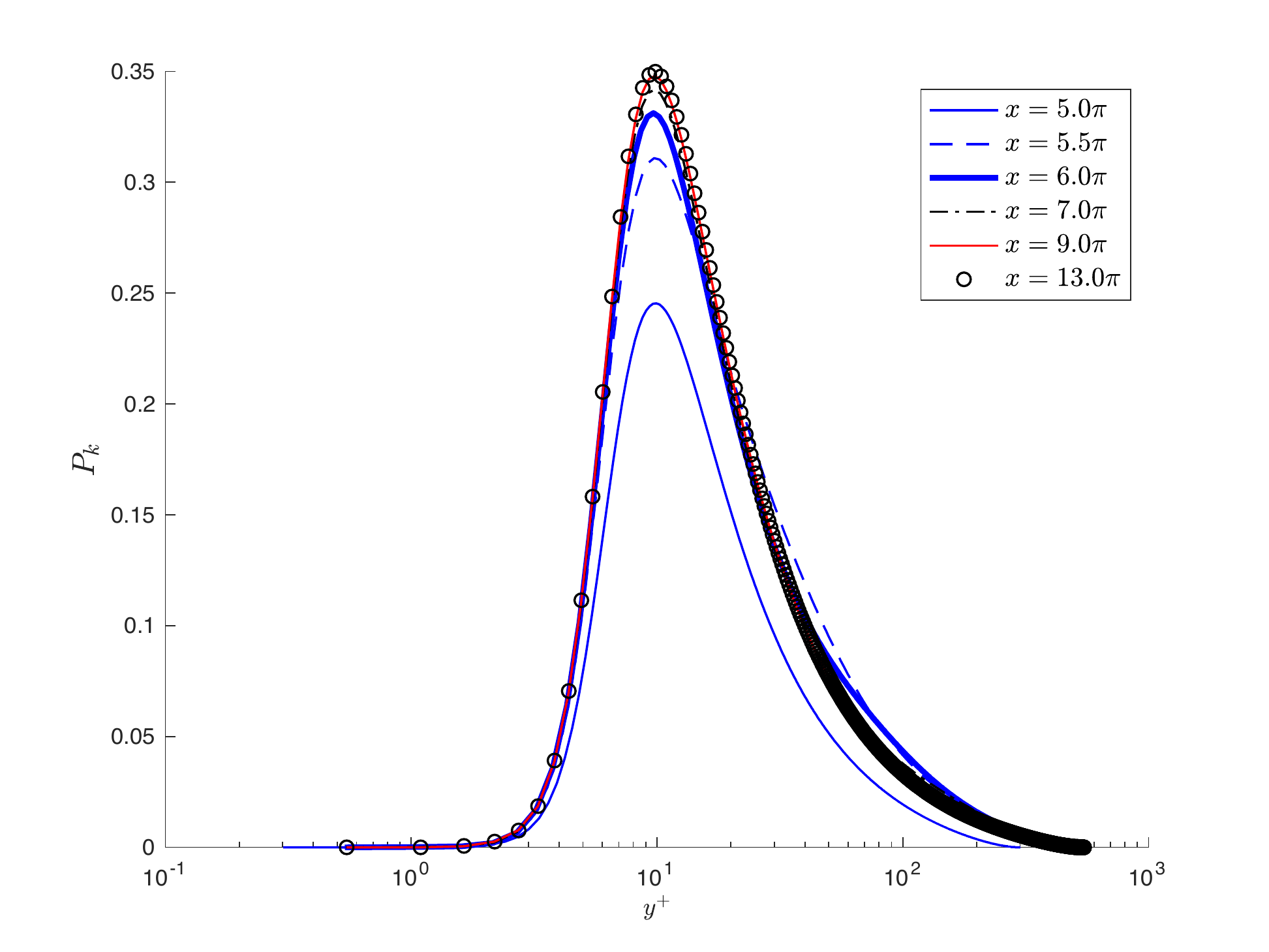}
\caption{The development of the production for the turbulent kinetic energy obtained using the standard \(k-\tau\) model available in Nek5000, i.e. without considering the correction proposed in Eq.~\ref{eq:correction}}
\label{fig:prod_development_stndrd}
\end{figure}

Fig.~\ref{fig:prod_development_stndrd} presents the development of the production term without the correlation proposed by Eq.~\ref{eq:correction}. In this plot, the production is normalized through \(u_\tau^4/\nu\), with values for the friction velocity \(u_\tau\) and \(\nu\) from Region I, which allow us to visualize absolute variations in this term.

\begin{figure}[htb]
\centering 
\includegraphics[width=\linewidth]{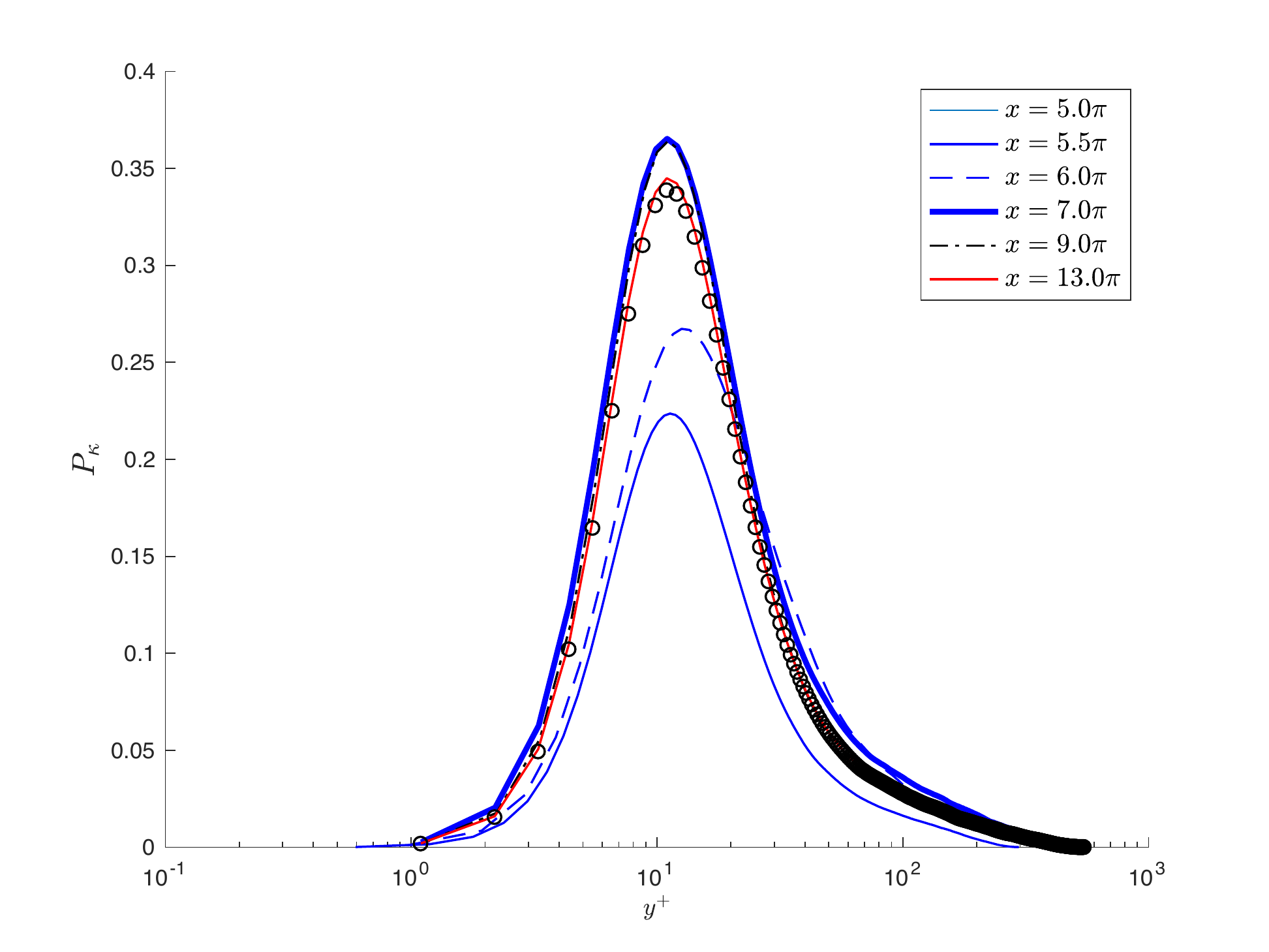}
\caption{The development of the production for the turbulent kinetic energy obtained via DNS~\cite{leite2021}}
\label{fig:prod_development_DNS}
\end{figure}

Fig.~\ref{fig:prod_development_DNS}, obtained from Ref.~\cite{leite2021}, is a similar plot to Fig.~\ref{fig:prod_development_stndrd}, but with the values of the production term obtained via DNS, as outlined in that reference. In contrast to what observed in the DNS results, the production development obtained using the standard \(k-\tau\) model does not retrieve all the effects due to spatially varying viscosity.

\begin{figure}[htb]
\centering 
\includegraphics[width=\linewidth]{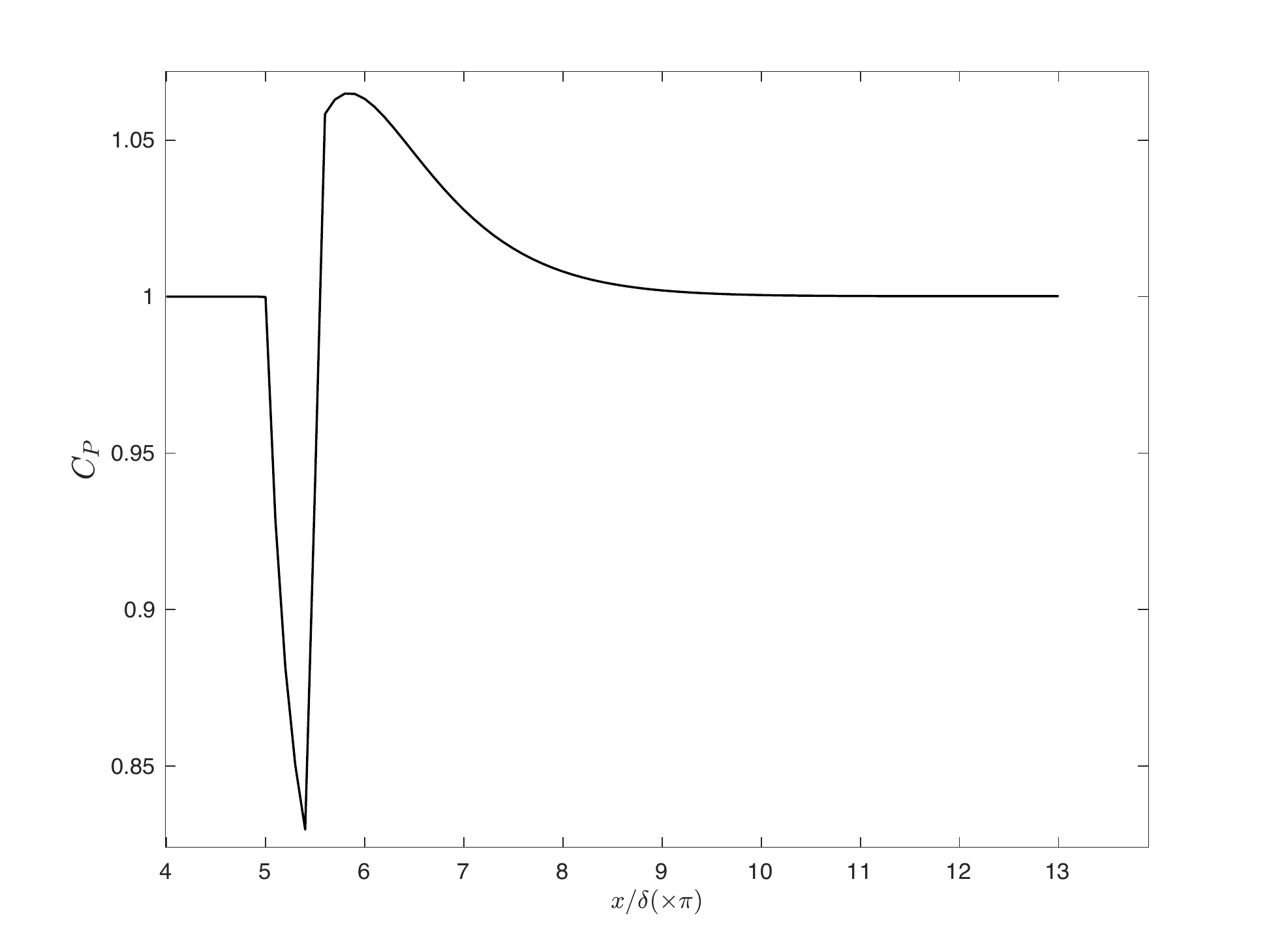}
\caption{The values for \(C_P\) obtained using Eq.~\ref{eq:correction} throughout the streamwise direction of the tested case}
\label{fig:cp_plot}
\end{figure}

\begin{figure}[htb]
\centering 
\includegraphics[width=\linewidth]{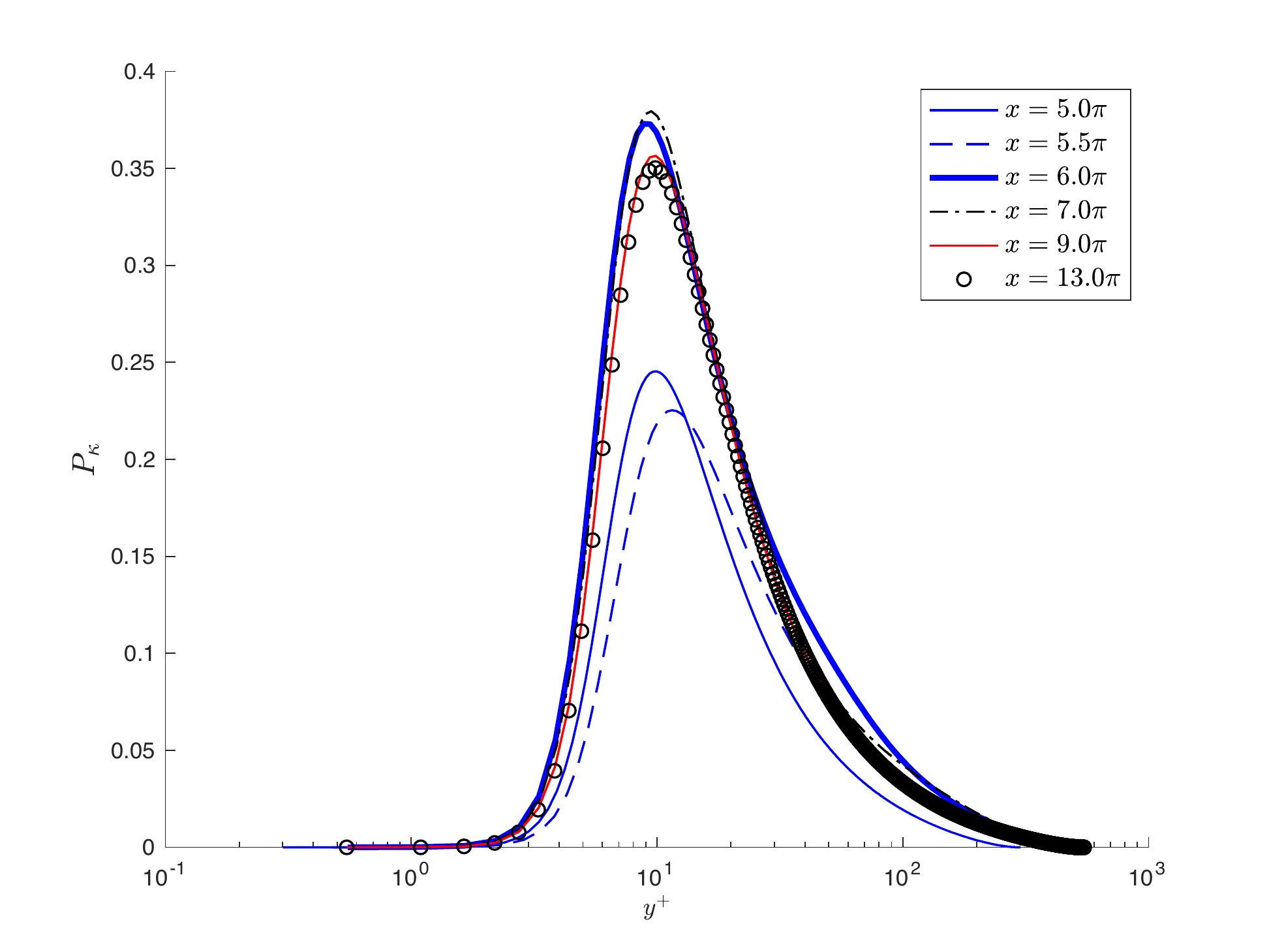}
\caption{The development of the production for the turbulent kinetic energy obtained using the corrected \(k-\tau\) model, this time considering the correction proposed in Eq.~\ref{eq:correction}}
\label{fig:prod_development_corrected}
\end{figure}

This way, we aim to reproduce the effects by employing the correction proposed by Eq.~\ref{eq:correction}. Fig.~\ref{fig:cp_plot} presents the values for \(C_P\) and then, Fig.~\ref{fig:prod_development_corrected} shows the value for \(P_\kappa\), i.e. the production considering the correction to account for spatially varying viscosity effects. By comparing Fig.~\ref{fig:prod_development_stndrd} and Fig.~\ref{fig:prod_development_corrected}, it is clear that we obtained an improvement in the representation of the production, likely because the modified results exhibits the turbulence bursts after the ramp region reported in Ref.~\cite{leite2021}.

\begin{figure}[htb]
\centering 
\includegraphics[width=\linewidth]{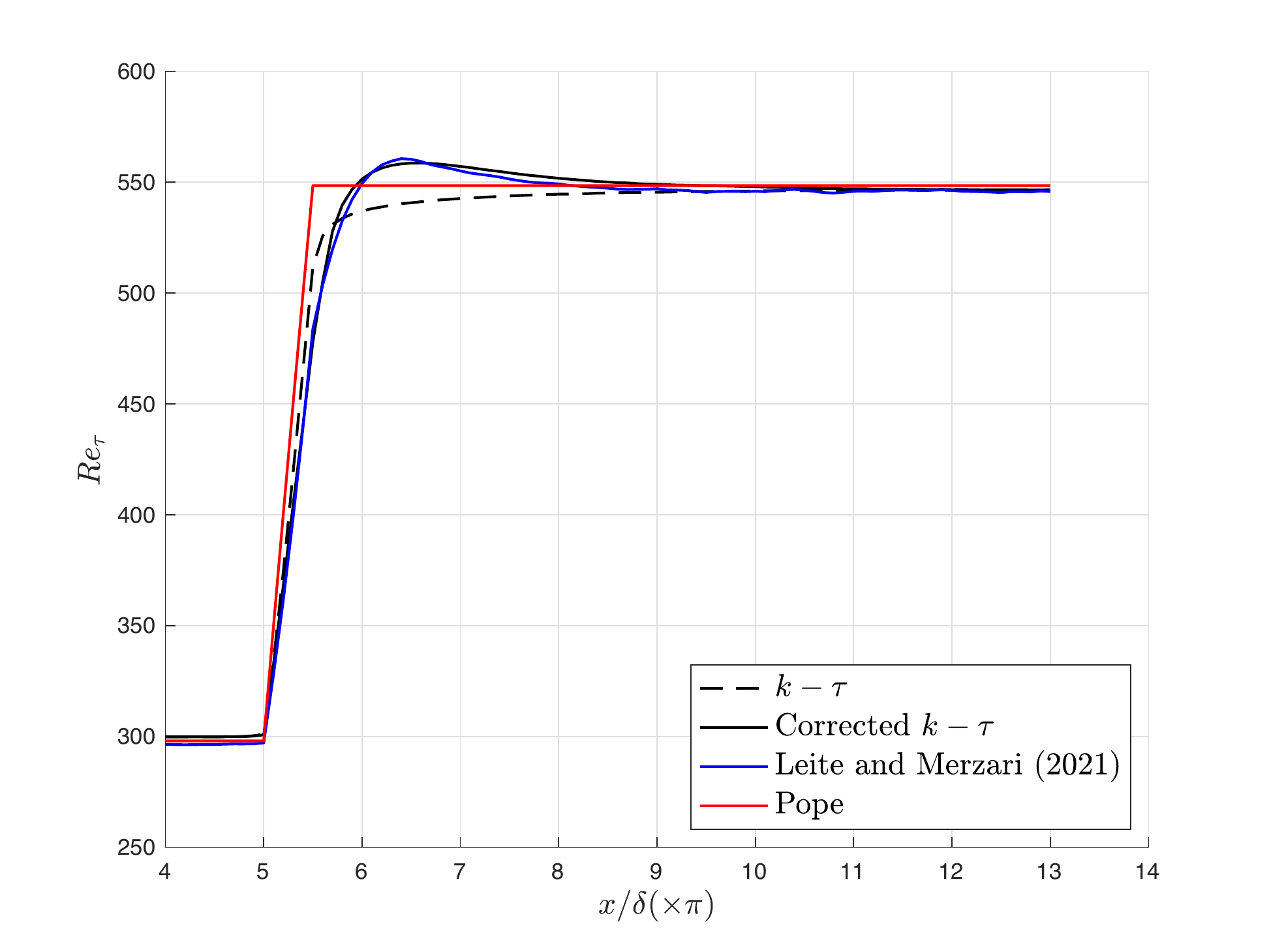}
\caption{The friction Reynolds number through the streamwise direction}
\label{fig:re_tau}
\end{figure}

Furthermore, we also compared other quantities to check the improvement provided  by  the proposed modifications. For example, Fig.~\ref{fig:re_tau} shows a plot comparing the local friction Reynolds number \(Re_\tau\). In this figure, besides showing the results for both the standard and the corrected \(k-\tau\) RANS models, the DNS results from the earlier work \cite{leite2021} and the correlation from Pope~\cite{pope2000} given by Eq.~\ref{eq:correlation_pope} are also presented. It is clear from this plot that the modification proposed allowed to better account for the effects reported in \cite{leite2021} caused by varying viscosity. With the correction, the delay in the ramp region can be simulated as turbulence does not respond instantaneously to the change of viscosity. At the same time, we are able to simulate the turbulence bursts that takes place right after the ramp region.

\begin{figure}[htb]
\centering 
\includegraphics[width=\linewidth]{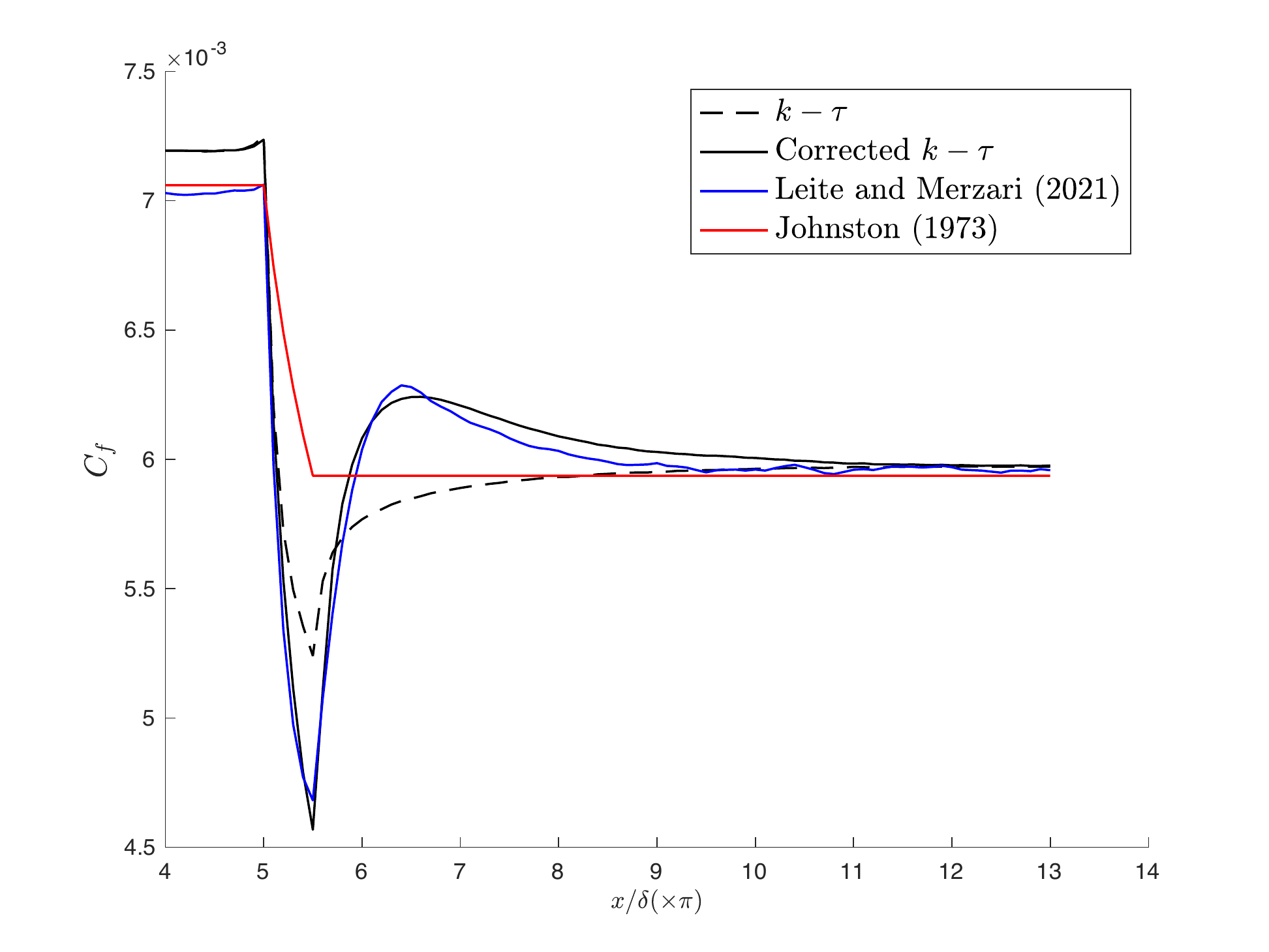}
\caption{The friction coefficient through the streamwise direction}
\label{fig:drag_coef}
\end{figure}

The improvement is also present when analyzing the friction coefficient \(C_f\) in Fig.~\ref{fig:drag_coef}. In this figure, the skin friction coefficient is calculated as \(C_f=\tau_w/(\rho U^2/2)\) using the simulation results, both via RANS and DNS. Additionally, the Johnston's correlation valid for fully developed flows, \(C_f=0.0706/Re^{0.25}\)~\cite{johnston1973} , is also presented for reference. Similarly to the plot for the friction Reynolds number, we can clearly see the improvement in the results when employing the proposed modification.

From Fig.~\ref{fig:drag_coef}, we report the standard deviation of the discrepancy between $C_f$ obtained via the standard version of $k-\tau$ and the DNS result from~\cite{leite2021} is $\sigma_{C_f}=2.0 \: E-04$, whereas the same value considering the modified version is of $\sigma_{C_f}^*=6.0 \: E-05$. These parameters serves to measure the agreement of each one of these two models with the DNS results. This way, we can conclude that the proposed model was able to improve by one order of magnitude the results for $C_f$ in the tested case.

\section*{CONCLUSIONS}

In the present paper, a modification in the standard $k-\tau$ model has been proposed in order to account for varying viscosity effects identified and characterized in Ref~\cite{leite2021}. The proposed modification consists on multiplying the original production term $P_k$ by a correction factor $C_P$, given by Eq.~\ref{eq:correction}.

The proposed correction is based on the friction Reynolds values yielded by a convolution function provided by Eqs.~\ref{eq:convolution1}-\ref{eq:convolution3}. This set of equations was developed in Ref~\cite{leite2021} to serve as a tool to model the effects of spatially varying the viscosity. It is clear that the convolution function plays an important role when accounting for the correction needed. Thus, it should be stressed that the proposed modified $k-\tau$ does not feature a general formulation, once the parameters from Eqs.~\ref{eq:convolution1}-\ref{eq:convolution3} may change depending on the inclination and the length of the imposed ramp. Finally, the model needs to be further tested and developed in order to obtain a more general formulation. Nevertheless, the strategy of accounting varying viscosity effects by correcting the production term proved to work well for the tested case.

\bibliographystyle{asmems4}

\bibliography{asme2e}
\appendix

\end{document}